\documentclass[useAMS,usenatbib]{mn2e}
\usepackage{epsfig,amsmath,amssymb,amsfonts,mathrsfs,latexsym,graphicx}
\include{journals}
\newcommand{\eg}{e.g.\ }
\newcommand{\ie}{i.e.\ }
\newcommand{\Msun}{{\rm M}_{\odot}}

\newcommand{\Lsun}{L_{\odot}}
\newcommand{\kms}{km~s$^{-1}$}
\newcommand{\ergs}{erg s$^{-1}$}

\newcommand{\OI}{O~{\sc i}}

\newcommand{\SiII}{Si~{\sc ii}}

\newcommand{\CaII}{Ca~{\sc ii}}

\newcommand{\FeII}{Fe~{\sc ii}}

\newcommand{\Fefs}{$^{56}$Fe}
\newcommand{\Cofs}{$^{56}$Co}
\newcommand{\Nifs}{$^{56}$Ni}

\newcommand{\Mej}{$M_{\rm ej}$}
\newcommand{\KE}{$E_{\rm kin}$}

\newcommand{\vph}{v_{\rm ph}}

\newcommand{\apj}{ApJ}
\newcommand{\apjl}{ApJL}

\newcommand{\aj}{AJ}
\newcommand{\aap}{A\&A}
\newcommand{\mnras}{MNRAS}
\newcommand{\nat}{Nat}
\newcommand{\pasp}{PASP}

\def\gsim{\mathrel{\rlap{\lower 4pt \hbox{\hskip 1pt $\sim$}}\raise 1pt \hbox {$>$}}}
\def\lsim{\mathrel{\rlap{\lower 4pt \hbox{\hskip 1pt $\sim$}}\raise 1pt \hbox {$<$}}}

\title[2010ah]{The very energetic, broad-lined type Ic Supernova 2010ah (PTF10bzf) in
the context of GRB/SNe}  

\author[P.A. Mazzali et al.]{
\parbox[t]{\textwidth}
{Paolo A. Mazzali$^{1,2,3}$\thanks{E-mail: mazzali@mpa-garching.mpg.de},  
Emma S. Walker$^{4,5}$, 
Elena Pian$^{5,6}$, 
Masaomi Tanaka$^7$, \\
Alessandra Corsi$^8$, 
Takashi Hattori$^9$,
and Avishay Gal-Yam$^{10}$}
\vspace*{6pt}\\
$^1$INAF-Osservatorio Astronomico, vicolo dell'Osservatorio, 5, I-35122 
Padova, Italy \\
$^2$Astrophysics Research Institute, Liverpool John Moores University,
Liverpool, UK \\
$^3$Max-Planck Institut f\"ur Astrophysik, Karl-Schwarzschildstr. 1, D-85748 
Garching, Germany \\
$^4$Yale University, New Haven, CT, USA \\
$^5$Scuola Normale Superiore, Pisa, Italy \\
$^6$INAF-IASF Bologna, via Gobetti 101, 40129 Bologna, Italy \\
$^7$National Astronomical Observatory, Mitaka, Tokyo, Japan \\
$^8$George Washington University, Washington, D.C., USA \\
$^9$Subaru Telescope, National Astronomical Observatory of Japan, Hilo, HI, USA \\
$^{10}$Weizmann Institute of Science, Rehovot, Israel }

\begin{document}

\date{Accepted ... Received ...; in original form ...}

\pagerange{\pageref{firstpage}--\pageref{lastpage}} \pubyear{2011}

\maketitle

\label{firstpage}


\begin{abstract}
SN\,2010ah, a very broad-lined type Ic Supernova (SN) discovered by the Palomar
Transient Factory, was interesting because of its relatively high luminosity and
the high velocity of the absorption lines, which was comparable to that of
Gamma-ray Burst (GRB)/SNe, suggesting a high explosion kinetic energy. However,
no GRB was detected in association with the SN. Here, the properties of
SN\,2010ah are determined with higher accuracy than previous studies through 
modelling. New Subaru telescope photometry is presented. A bolometric light
curve is constructed taking advantage of the spectral similarity with
SN\,1998bw.  Radiation transport tools are used to reproduce the spectra and the
light curve. The results thus obtained regarding ejecta mass, composition and
kinetic energy are then used to compute a synthetic light curve. This is in
reasonable agreement with the early bolometric light curve of SN\,2010ah, but a
high abundance of \Nifs\ at high velocity is required to reproduce the early
rise, while a dense inner core must be used to reproduce the slow decline at
late phases. The high-velocity \Nifs\ cannot have been located on our line of
sight, which may be indirect evidence for an off-axis, aspherical explosion. 
The main properties of SN\,2010ah are: ejected mass \Mej\:$\approx 3 \Msun$;
kinetic energy \KE\: $\approx 10^{52}$\,erg, M(\Nifs)\:$ \approx 0.25 \Msun$.
The mass located at $v \gsim 0.1$\,c is $\sim 0.2 \Msun$. Although these values,
in particular the \KE, are quite large for a SN\,Ic, they are all smaller 
(especially \Mej) than those typical of GRB/SNe. This confirms the tendency for 
these quantities to correlate, and suggests that there are minimum requirements 
for a GRB/SN, which SN\,2010ah may not meet although it comes quite close. 
Depending on whether a neutron star or a black hole was formed following core 
collapse, SN\,2010ah was the explosion of a CO core of $\sim 5-6 \Msun$, 
pointing to a progenitor mass of $\sim 24 - 28 \Msun$.
\end{abstract}



\section{Introduction}

Type Ib/c supernovae are perhaps the most diverse subtype of SNe. Their
spectroscopic definition \citep[no H (type Ib), no He (type Ic), weak Si,
see][]{Fil97} means that this is the class in which almost all ``strange" SNe
are confined. SNe\,Ib/c include all explosions of massive stars that have lost
their outer envelopes. The diversity of possible progenitors and evolutionary
histories is reflected in the huge variety of properties of SNe\,Ib/c, which
range from low (``normal"?) luminosity and kinetic energy \citep[\eg
SN\,1994I,][]{Richmond94I,Iwamoto94I,Sauer06} to broad-lined SNe\,Ic \citep[\eg
SN\,2002ap,][]{Mazz02ap,Foley03,GalYam02}, to XRF-connected SNe\,Ib \citep[\eg
SN\,2008D,][]{sod08, Mazz08D, modjaz08} and SNe\,Ic \citep[\eg
SN\,2006aj,][]{Pian06}, to luminous and energetic GRB/SNe\,Ic (SN\,1998bw,
\citet{gal98}; SN\,2003dh, \citet{stanek03,hjorth03,Matheson03}; SN\,2003lw,
\citet{Malesani04}).   The SN\,IIb subtype, where a small fraction of the H
envelope survives, showing up in the spectrum but leaving practically no imprint
on the light curve \citep{Haching12}, also belongs to the same physical group.
SNe\,IIb are also known to range from low \citep[\eg SN\,1993J,][]{Fil93J} to
high energy events \citep[\eg SN\,2003bg,][]{Hamuy03gb}. Binarity is suspected
to play a major role in the evolution of their progenitors \citep{nomoto_bin,
maund93J, crockett08ax, Arcavi11, vandyk11dh, maund11dh}. Some very bright
SNe\,Ic are possibly linked to the pair instability phenomenon in very massive
stars \citep[\eg the SN\,Ic 2007bi,][]{GalYam07bi}. Among the members of this
class are also very luminous SNe whose origin is still debated (\eg SN\,2005ap,
\citet{quimby07,quimby11,GalYam12}, and references therein) or PS1-10afx,
\citep{Chornock13}.  Finally, at the opposite end of the distribution, very dim
SNe\,Ib such as 2005E and other ``Ca-rich SNe" \citep{perets09,kasliwal12} may
be the result of the explosive ejection of a shell from the surface of a
He-accreting white dwarf. All these SNe are also formally members of the
SN\,Ib/c class, but their physical origin can be very different. 

Still, the connection between SNe\,Ib/c, massive stars and GRBs is one of the
most important reasons to study these stellar explosions. While extremely massive
stars (M$ \geq 80 \Msun$) are believed to explode via the Pair Instability
mechanism, which leaves no remnant, stars more massive than $\sim 25 \Msun$
should collapse to form black holes. This mechanism is thought to be responsible
for the emission of long-soft GRBs \citep{woosley93}. Although the details of the
energy extraction are not yet understood, there is sufficient energy to power a
relativistic jet \citep{McFW99,Bromberg11}.  In fact, all well observed GRB/SNe
are broad-lined SNe\,Ic, and all have been inferred to have large ejected mass
and kinetic energy \citep{iwa98, Mazz03dh, Mazz03lw}. These quantities are
derived from modelling the light curves and spectra of these SNe, and the results
consistently yield ejected masses of $\sim 10 \Msun$, kinetic energies of several
$10^{52}$\,erg, \Nifs\ masses of $\sim 0.5 \Msun$, which have been used to infer
progenitor star masses of $35 - 50 \Msun$. 

On the other hand, X-ray Flash/SNe like SN\,2006aj seem to come from stars of
lower mass ($\sim 20 \Msun$), suggesting that energy injection from a magnetar
is the mechanism which originated the XRF and energised the explosion
\citep{Mazz06aj}. 

There are various cases of broad-lined SNe\,Ic whose properties are intermediate
between those of GRB/SNe and those of XRF/SNe and are apparently not accompanied
by a relativistic outflow (\eg SN\,1997ef \citep{Mazz97ef}; SN\,1997dq
\citep{Mazz97dq}; SN\,2002ap \citep{Mazz02ap}). This may suggest that some of
the properties of these SNe were not sufficient for the generation of a GRB.
Alternatively, they may all have off-axis GRBs, but this seems to be ruled out
by radio observations \citep{sod06}. Studies of the emission line profiles in
the late, nebular phase \citep{Mazz97dq,Mazz06ajneb,Mazz07gr} do not suggest the
off-axis option for most of these SNe, although they clearly do so for some of
them \citep{Mazz03jd,Tanaka08}. Spectropolarimetry, which is available only in
very few cases, also suggests that SNe\,Ic are aspherical 
\citep[\eg][]{Kawabata02,Kawabata03,Gorosabel06,Gorosabel10}, and that more
energetic SNe are more aspherical than lower-energy ones \citep{Tanaka12}.

Clearly, in order to improve our understanding of these explosions, we need to
extend the observed sample, cover a broader range of SN properties. A
particularly interesting domain is SNe\,Ib/c with properties straddling those of
GRB- and non-GRB/SNe. The presence of a high-energy transient is certainly one
of the dimensions that determine SN\,Ib/c properties, although not the only
one.  One of the few SNe that falls in this category is SN\,2010ah/PTF10bzf, a
bright, broad-lined SN\,Ic without a GRB, for which extremely high velocities
were inferred \citep{corsi11}. In this paper we model the spectra and light
curve of SN\,2010ah with our Montecarlo radiation transport codes and determine
its properties. Our work is based on the data presented by \citet{corsi11}, but
we add very accurate photometry obtained with the Subaru Telescope.

The paper is organised as follows. In section 2 we recap the properties of
SN\,2010ah. In section 3 we present the new Subaru photometry. In section 4 we
discuss how we use the spectral similarity to SN\,1998bw to build a bolometric
light curve for SN\,2010ah given the lack of multicolour photometry. The light
curve of SN\,2010ah is then compared to those of other hypernovae (HN). In
section 5 we briefly recap our modelling method and present spectral models for
SN\,2010ah. In Section 6 the synthetic light curves computed based on the
spectral modelling results are presented and compared to the bolometric light
curve of SN\,2010ah. Finally, Section 7 contains a discussion and conclusions.

\section{PTF10bzf = SN\,2010ah: observational properties}

The bright, broad-lined SN\,Ic SN\,2010ah was discovered by the Palomar
Transient Factory \citep[PTF,][]{law09,rau09} on 23 Feb 2010 in a galaxy at a
redshift $z = 0.0498$ \citep{corsi11}. It reached $r = 18.3$\,mag on 2 Mar 2010.
At a distance modulus of $36.7$\,mag, this corresponds to an absolute magnitude
$M_R \approx -18.4$\,mag. This makes SN\,2010ah a luminous SN, but not a
record-breaker. Data from the PTF follow-up were published in \citet{corsi11}.
The SN was classified as a Type Ic from spectra obtained at Gemini-N and Keck
Obs, which show very broad lines.  Velocities in excess of $0.1 c$ were measured
in \CaII\ and \SiII\ lines \citep[][fig. 5]{corsi11}. In fact, absorption lines
in SN\,2010ah are among the broadest ever, rivalling with GRB/SNe such as
1998bw. \citet{corsi11} suggested that the SN is intermediate in spectral
properties between the GRB/SN\,1998bw \citep{patat01} and the non-GRB SN\,1997ef
\citep{Mazz97ef}. Actually, SN\,1998bw is the closest match, as can be seen from
figs. 2 and 3 of \citet{corsi11}. However, no X-ray counterpart was detected for
SN\,2010ah. 

The striking spectral similarity to SN\,1998bw and the lack of an associated
$\gamma$- or X-ray counterpart make SN\,2010ah an interesting event. The
$R$-band luminosity was comparable to that of other broad-lined SNe\,Ic without
a GRB, like SN\,2003jd or SN\,2009bb, for both of which an off-axis GRB has been
suggested \citep[][respectively]{Mazz03jd,sod09bb}. However, SNe\,2003jd and
2009bb resemble one another spectroscopically and look much more like the
XRF/SN\,2006aj \citep[][figs. 11, 12]{pignata11} than SN\,2010ah or SN\,1998bw.

In Figure 1 we compare the two available spectra of SN\,2010ah with the spectra
of SNe\,1998bw and 2002ap that look most similar based on an eye estimate of the
available ones.  The overall similarity is apparent: all three SNe show extreme
line blending. In particular, \CaII\;IR and \OI\;7774\,\AA\ are blended, which
requires sufficient ejecta with $v \gsim 0.1 c$. In the blue, most
pseudo-emission peaks (which actually correspond to regions of low line opacity
where photons can escape) are obliterated because of line blocking by
high-velocity material (Fe, Cr, Ti). This effect is more apparent in SN\,1998bw
than in either SN\,2002ap or 2010ah, indicating a higher metal content at high
velocity. On the other hand, SN\,2010ah seems to have the highest line velocity
of the three SNe \citep{corsi11}. This is shown in particular by the blueshift
of the \SiII\;6355\,\AA\ absorption in the 7 Mar spectrum, which has $v \approx
18000$\,\kms. Considering that the epoch of the spectrum is at least 12 days
after explosion \citep{corsi11}, this velocity is exceptionally high, and
similar only to GRB/SNe \citep{Pian06,Bufano12}. 

The spectra of SN\,2010ah and SN\,2002ap used for the comparison are closer to
one another in time than those of SN\,1998bw.  Based on the epochs of the
spectra it can be roughly estimated that the time evolution of SN\,2010ah is
faster than that of SN\,1998bw by about a factor 1.3. In Section 4 we show how
this is in fact confirmed by the evolution of the light curves.

\begin{figure}
\includegraphics[width=89mm]{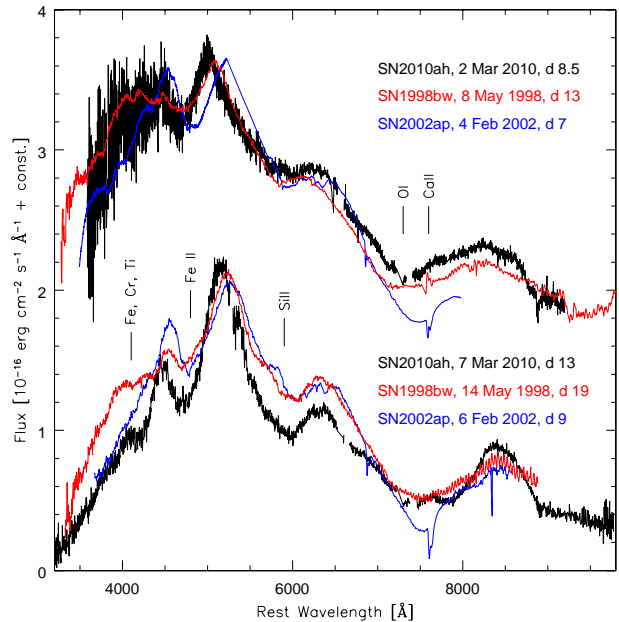}
\caption{Comparison of the spectra of SNe 2010ah, 1998bw \citep{patat01} and
2002ap (WHT service programmes) at similar epochs relative to the respective
light curve evolution. Ions responsible for the strongest spectral features are
marked.}
\label{comp}
\end{figure}

\section{New Subaru photometry and the light curves of SN\,2010ah}

The PTF follow-up campaign collected a good $r$-band light curve of SN\,2010ah,
but other bands were poorly sampled \citep{corsi11}. New Subaru data presented
here improve the coverage in the post-maximum phase and make the calculation of
a bolometric light curve more reliable. 

Imaging observations of SN 2010ah in the $B$, $V$, and $R$-bands were performed
under non-photometric conditions with the Subaru Telescope equipped with the
Faint Object Camera and Spectrograph \citep[FOCAS,][]{kashikawa02} on 2010 Mar
19.25, 20.25, 21.25, and 22.24 UT (MJD = 55274.25, 55275.25, 55276.25, and
55277.24, respectively). The exposure time was 2 sec for each image. The images
were bias-subtracted and flat-fielded. Photometry was performed relative to 5
stars in the field for the $R$ and $V$-band images and to 4 stars for the
$B$-band images. We used an aperture of 1.5 times the full width at half-maximum
(FWHM) of the point-spread function (PSF) of each image, estimated as the mean
of the FWHM measured over the references stars and SN\,2010ah. The magnitudes of
the reference stars, originally in the SDSS system, were calibrated to $BVR$
using the conversions described in \citet{Jordi06}. Color corrections were
applied to each image assuming that the measurement using the filter at the
adjacent longer wavelength contains the information required to perform the
correction. Zero-points and colour term coefficients were derived performing a
linear fit on the reference stars for each image, and then used to derive the SN
photometry for the corresponding image. The $B-V$, $V-R$, and $R-I$ colours of
SN\,2010ah were derived from WISE photometry \citep{corsi11} at epochs close to
those of the Subaru images. Since only 2 reference stars could be identified in
the B-band image of Mar 19.25, the zero-point and colour term could not be
determined, and a magnitude from this image could not be derived.

\begin{figure}
\includegraphics[width=89mm]{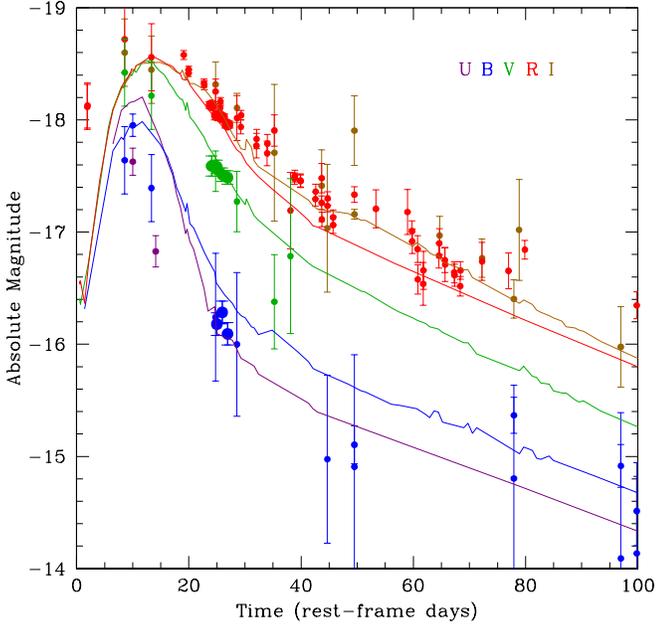}
\caption{The UBVRI light curves of SN\,2010ah. The light curves, dereddened with
$E_{B-V} = 0.012$, include the data presented in \citet{corsi11} and the new
Subaru photometry (larger dots).  The lines show the light curves of SN\,1998bw,
dereddened with $E_{B-V} = 0.016$, dimmed by 0.5 mags and stretched in time by a
factor $1.3^{-1}$.  The light curves of SN\,2010ah approximately match those of
SN\,1998bw.  } 
\label{ubvri}
\end{figure}

Figure 2 shows the multiband photometry of SN\,2010ah from \citet{corsi11} and
the new Subaru data, highlighted as large circles.  In order to build this
figure, the photometric data originally acquired with filters different from the
Johnson-Cousins system in the PTF campaign have been reduced to this system as
follows. For the P48 data, $R = R_{Mould} - 0.1$ was assumed.  The P60 $griz$
magnitudes were transformed to $BRI$ using colours from simultaneous
observations and normalisations as in \citet{Fukugita96}. Lacking spectroscopic
information for SN\,2010ah after maximum and based on the spectral similarity
with SN\,1998bw (see Figure 1), the $g,r,i,z$ and Mould-$R$ magnitudes of
SN\,2010ah have been converted to the $BVRI$ system and k-corrected following
\citet{Hamuy93} and \citet{oke68}, using as templates the spectra of SN\,1998bw
at phases comparable to those of SN\,2010ah, after application of a time-folding
factor of 1.3.  The UVOT U-band filters were transformed to $U$ magnitudes using
the zero-points of \citet{poole08} and a $\lambda^3$ power-law for the UV
spectrum. Finally, synthetic magnitudes were extracted from the Gemini and Keck
spectra. As these were both taken with the slit positioned near the parallactic
angle, we assumed that in both cases flux calibration is reasonable across the
whole spectrum. When we measured the synthetic photometry we conservatively
assigned an uncertainty of $\pm 0.3$\,mag to both points, as in \citet{corsi11}.
The error on the first 2 $R$-band points has been increased to 0.2\,mags because
the k-correction for those points is based on extrapolation from a SN\,1998bw
spectrum whose epoch is 3.5 days later in the time frame of SN\,2010ah.  At
early epochs the spectral shape may change significantly over this time
interval.

A fiducial time of explosion was selected. Since the explosion date is
constrained by an upper limit on 19 Feb 2010, 4 days before the first PTF
detection \citep[][fig. 2]{corsi11}, we assumed that the explosion occurred 2
days before first detection. 

In Fig. 2 all magnitudes have been dereddened with $E_{B-V} = 0.012$
\citep{Schlegel98}, adopting $R_V = 3.1$ and the Galactic extinction curve of
\citet{cardelli89}.  The comparison of the rest-frame light curves of SN\,2010ah
with the much better sampled ones of SN\,1998bw in the same filters, and
especially in the $R$-band, where SN\,2010ah has been most intensively observed,
confirms that the time evolution of SN\,2010ah is faster than that of SN\,1998bw
by a factor of $\sim 1.3$.  If the rest-frame monochromatic light curves of
SN\,1998bw are divided by this factor they  match those of SN\,2010ah reasonably
well (Fig. 2), although bluer bands in SN\,2010ah seem to evolve slightly faster
than redder bands with respect to SN\,1998bw.  In the $U$-band, the two points
available for SN\,2010ah are below the SN\,1998bw template, suggesting a more
rapid decay in this band, and perhaps some modest local absorption.

\section{SN\,2010ah: the bolometric light curve}

In order to build a successful model for SN\,2010ah it is necessary to construct
a bolometric light curve. This is essential for the estimate of the mass of
\Nifs\ synthesised in the explosion and as an input to our spectrum-synthesis
code. Since PTF photometry was obtained almost exclusively in the $r$-band, it
is difficult to construct a well sampled bolometric light curve. At the same
time it would be dangerous to use $r$ as a proxy for bolometric because the
bolometric correction from $r$ may be different from SN to SN. 

In order to compute the bolometric light curve of SN\,2010ah we made use of the
spectroscopic similarity between SN\,2010ah and SN\,1998bw and of the similar
temporal evolution after rescaling in time by a factor $1.3^{-1}$ \citep[Figure
2, see also][]{Deng05}. We therefore adopted for SN\,2010ah the bolometric
correction from $r$ obtained for SN\,1998bw \citep{patat01} at corresponding
epochs (stretched in time by $1.3^{-1}$). The resulting bolometric light curve
is shown in Fig. 3. The epoch and brightness of the peak are not very well
determined, but the good coverage of the decline phase allows a model to be
developed.

As a verification of this method, for the two epochs when spectra are available
we integrated the flux under the spectra, added 20\% to account for flux outside
the optical range taking SN\,1998bw as an example \citep[\eg][]{Melandri12}, and
allowed for a conservatively large error to account for uncertain flux
calibration of the spectra. 

Also, again as a verification, a bolometric magnitude was computed directly from
the monochromatic data when sufficient coverage of the spectrum was available.
This includes the epochs of the Subaru photometry, when quasi-simultaneous WISE
$I$-band data are also available \citep{corsi11}, and an epoch near 50 days,
when $b,r$ and $i$ photometry is available. A correction of 20\% was again added
for flux outside the optical bands. These points are highlighted in Figures 3
and 4.

\begin{figure}
\includegraphics[width=89mm]{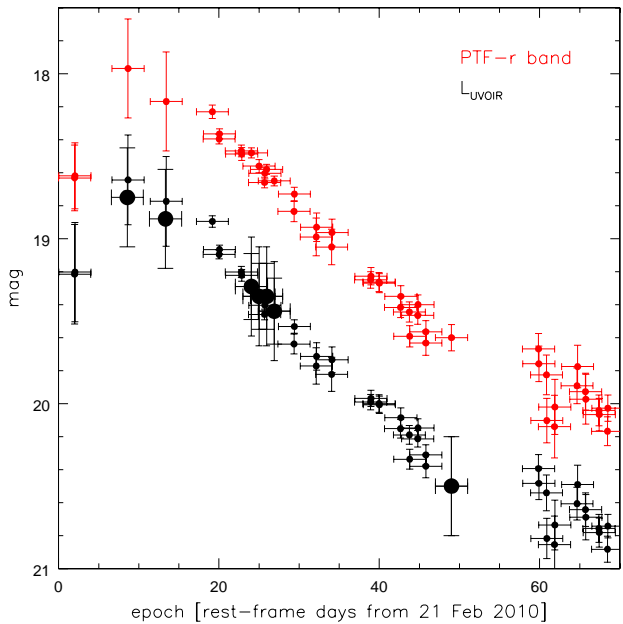}
\caption{The PTF-r band light curve of SN\,2010ah (red) and the bolometric
light curve built from the photometry and correction with respect to 
SN\,1998bw (black). Larger black points indicate estimates obtained directly
from the spectra (near maximum) and from multi-colour photometry.}  
\label{lc}
\end{figure}

\begin{figure}
\includegraphics[width=89mm]{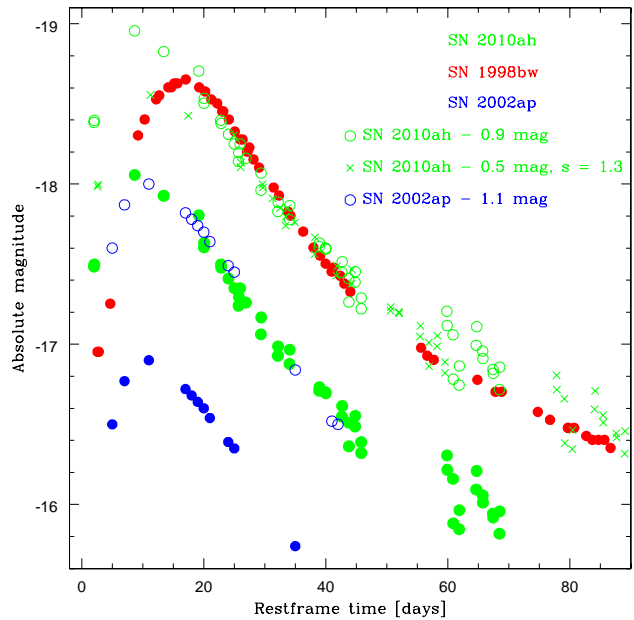}
\caption{Comparison of bolometric light curves of different SNe\,Ic: 
Filled dots show the bolometric light curve of SN\,2010ah, 1998bw and 2002ap.
Empty green circles show the light curve of SN\,2010ah shifted upwards by 0.9
mags. This is a good match to SN\,1998bw in the declining phase. 
Green crosses show the light curve of SN\,2010ah shifted upwards by 0.5 mags 
and stretched in time by a factor $(1.3)$. This is a good match to SN\,1998bw
except the earliest epochs.
Empty blue dots show the light curve of SN\,2002ap shifted upwards by 1.1 mags.
This is a reasonable match to SN\,2010ah.} 
\label{lccomp}
\end{figure}

\subsection{SN properties derived from the light curve} 

A useful first step towards establishing a model for SN\,2010ah is to derive
basic information from the light curve and the spectra. In the spirit of
\citet{Arnett82} this requires measuring the width of the light curve and a
characteristic ejecta velocity from the spectra. These in turn can be
transformed into first guesses for the ejected mass and kinetic energy. In Fig.
4 we compare the bolometric light curve of SNe\,2010ah, 1998bw and 2002ap, all
of which look similar spectroscopically. The light curve of  SN\,2010ah is very
similar in shape to that of SN\,2002ap, showing a fast rise to peak, which is
reached within $\sim 10$ days.  The light curve of SN\,2010ah is $\sim 5$\%
broader than that of SN\,2002ap, but significantly narrower ($\sim 25$\%) than
that of SN\,1998bw.  The characteristic velocity (at maximum) of SN\,2010ah can
be derived from fig. 5 of \citet{corsi11}. 

In order to obtain a rough estimate of the properties of SN\,2010ah  we can 
apply the well-known relations between ejected mass and velocity on the one hand
and kinetic energy and light curve shape on the other \citep{Arnett82}: 
\begin{eqnarray}
E_{\rm kin} & \propto & M_{\rm ej} v^2 \\
  \tau_{LC} & \propto & \frac {\kappa^{1/2} M_{\rm ej}^{3/4}}{E_{\rm kin}^{1/4}},
\end{eqnarray}
where \KE\ is the kinetic energy of the ejecta, \Mej\ the ejected mass, $\kappa$
the optical opacity, $\tau_{LC}$ the characteristic width of the light curve.
Adopting the commonly made assumption that $\kappa$ is constant in all
SNe\,Ib/c, simple scaling relations can be obtained:

\begin{eqnarray}
\frac{M_1}{M_2} & \propto & \left( \frac{\tau_1}{\tau_2}\right)^2 \frac{v_1}{v_2} \\
\frac{E_1}{E_2} & \propto & \left( \frac{\tau_1}{\tau_2}\right)^2 \left( \frac{v_1}{v_2}\right)^3
\end{eqnarray}

The results that are obtained from these relations depend sensitively on the
choice of $v$. This is a characteristic ejecta velocity, and in itself it is not
a very well defined quantity. Measuring line velocity and using that value has a
number of shortcomings: 1) the velocity changes (decreases) with time; 2) the
velocity at which lines form can be different (higher) than the momentary
photosphere; 3) different lines may form at different velocities because of
ionization or optical depth effects. 

Adopting a velocity used for modelling (as for example was done for the SNe
shown in fig.5 of \citet{corsi11}, with the exception of SN\,2010ah itself)
eliminates at least some of these problems. Still, a single velocity does not
represent the actual velocity distribution of the ejected matter. The velocity
changes rapidly with time, and light curves have different shapes. 

Very different results can be obtained if velocities at the same SN physical age
(\ie days from explosion, which is also an uncertain quantity except for GRB/SNe
or for cases where shock breakout is detected) or at a particular {\em phase} of
the light curve (\eg maximum) are used, since different SNe peak at different
physical times from explosion. Let us show some examples of this.

Since the epoch of maximum of the light curve of SN\,2010ah is not well defined,
different epochs can be chosen. Additionally, only two spectra are available. We
use the velocity derived from the second spectrum (7 March 2010), 18000\, \kms.
\citet{corsi11} assigned to this spectrum an epoch of 11 days, based on the
first detection.  Corresponding velocities at this epoch can be obtained for
both SN\,1998bw (via interpolation) and SN\,2002ap (from direct modelling). All
values are shown in Table 1. The value of \KE\ changes somewhat if we start from
SN\,1998bw or from SN\,2002ap, reflecting the uncertainty in LC shape and time
of maximum for SN\,2010ah.  Alternatively, we can consider the non-detection
limit 4 days earlier, and assign to the March 7 spectrum a fiducial epoch of 13
days, as we do in the spectroscopic analysis below.  The velocities of
SNe\,1998bw and 2002ap change significantly, despite the small change in epoch,
and so do the results obtained for \Mej\ and \KE\ of SN\,2010ah. The value of
\KE, in particular, is now much higher.  Indicative results for SN\,2010ah are
\Mej$\: \sim 4-6 \: \Msun$, \KE$\: \sim (1.5-3) \: 10^{52}$\,erg, with large
uncertainties. Similarly uncertain results would be obtained if we used the
earlier spectrum as a reference for velocity. This shows that the rescaling
method should be applied with caution, especially if the data coverage is not
excellent. Furthermore, it is imperative that SNe are used that have very
similar spectral properties in order to avoid carrying incorrect information
(especially about \KE\ as derived from velocities) from the SN used as template
to the one which is being studied.


\begin{table}
\begin{center}
  \caption{Derived properties of SN\,2010ah (\Mej, \KE) with different 
  assumptions about the epoch and using different reference SNe, whose \Mej\ 
  and \KE\ are given as are their velocities at different reference epochs.}
  \begin{tabular}{ccccc}
  \hline
 SN &  $\tau$ & vel (11 d) & \Mej & \KE \\
    &   days  & \kms  & $\Msun$ & $10^{51}$ erg \\
 \hline
 1998bw & 17 & 22500 & 10   & 50 \\
 2002ap & 12 & 12000 &  2.5 &  4 \\
 2010ah & 13 & 18000 &  5   & 15 (using 98bw) \\
        & 13 & 18000 &  4   & 15 (using 02ap) \\
 \hline
 SN &  $\tau$ &  vel (13 d) & \Mej & \KE \\
    &   days  & \kms  & $\Msun$ & $10^{51}$ erg \\
 \hline
 1998bw & 17 & 19000 & 10   & 50 \\
 2002ap & 12 & 10000 &  2.5 &  4 \\
 2010ah & 13 & 18000 &  6   & 25 (using 98bw) \\
        & 13 & 18000 &  5   & 25 (using 02ap) \\
 \hline
 \end{tabular}
\end{center}
\end{table}

The other important quantity that can be derived by interpolation is the mass of
\Nifs. This can be done using as reference the peak of the light curve, or
alternatively the late phase. The use of different and independent methods makes
the estimate of the M(\Nifs) more reliable than those of \Mej\ or \KE. Fig. 4
shows the two different approaches. The temporal evolution of the bolometric
light curve of SN\,2010ah looks rather similar to that of SN\,2002ap. If the
bolometric light curve of SN\,2002ap is shifted up by 1.1 mag it tracks that of
SN\,2010ah very well, except for the first point, which suggests that SN\,2010ah
had a broader light curve. The early rise of the light curve of SN\,2010ah may
be due to mixing out of some \Nifs, possibly in an aspherical explosion like
that of SN\,1998bw \citep{Maeda02,Maeda06}. A shift of 1.1\,mags corresponds to
a factor 2.75 in luminosity. The mass of \Nifs\ for SN\,2002ap was determined
from modelling as $0.07\: \Msun$ \citep{Mazz02ap}. This yields for SN\,2010ah
M(\Nifs$)\; = 0.19\: \Msun$, consistent with the value derived by
\citet{corsi11}. 

If we use as reference SN\,1998bw, on the other hand, we can follow two
different approaches. The post-maximum LC of SN\,2010ah can be superposed on
that of SN\,1998bw if it is scaled up by 0.9\,mags. This is a factor of 2.3 in
luminosity. Adopting for SN\,1998bw a \Nifs\ mass of $0.41\: \Msun$
\citep{nakamura00,Mazz03lw}, a \Nifs\ mass of $0.19\: \Msun$ is obtained for
SN\,2010ah, consistent with the value obtained when rescaling from SN\,2002ap.
In fact, differences in the relative times are negligible at advanced epochs.
This is therefore the safest approach.

Alternatively, one can rescale the bolometric light curve of SN\,2010ah in flux
AND stretch it by a factor 1.3 to generate a light curve that looks like that of
SN\,1998bw. In this case the rescaling factor is only 1.6 (0.5 mags), and the
\Nifs\ mass for SN\,2010ah is $0.26\: \Msun$. This result is clearly
inconsistent. The different time evolution of the two light curves must be taken
into account: at early times the contribution of \Cofs\ can be ignored, and
shifting the peak from 17 days (SN\,1998bw) to 13 days (SN\,2010ah) leads to a
\Nifs\ mass for SN\,2010ah smaller by a factor of 0.63 with respect to the
inital estimate. Hence we recover M(\Nifs)$ = 0.16 \Msun$, which is reasonably
consistent with the estimate based on the light curve tail. We conclude that for
SN\,2010ah M(\Nifs)$= 0.18 \pm 0.02 \Msun$. A departure from this value can be
introduced by a) any \Nifs\ which is mixed to high velocities and is only
visible in the rising phase; b) some deeply hidden \Nifs\ which at peak does not
yet contribute to the luminosity and only affects the light curve at very late
times. This central \Nifs\ is known to be present in a high density inner region
in probably all hypernovae \citep{maeda03}. A signature of this deep \Nifs\ may
also be seen in SN\,2010ah after about day 40, when the LC begins to flatten,
although data unfortunately stop soon thereafter. Therefore, M(\Nifs)$ = 0.20
\Msun$ is probably a lower limit for the mass of \Nifs\ in SN\,2010ah.

\section{Spectral modelling}

Given the description above, only accurate spectral (and light curve) modelling
can yield reliable information on the properties of SNe\,Ib/c. For SN\,2010ah
only two spectra are available, so a very accurate result should not be
expected. However, since the two spectra are separated by several days modelling
may yield significant information. 

We used our SN Montecarlo spectrum synthesis code, which was first presented by
\citet{m&l93}. The code computes radiation transport in a SN ejecta starting
from an assumed luminosity emitted at a lower boundary as a black body. SN
ejecta are assumed to be in free expansion, so that $v = R/t$. Bound-free
processes are ignored, but electron scattering and line absorption are
considered. The latter is treated including the process of fluorescence, which
is essential for the formation of type I SN spectra \citep{l99,maz00}. Radiative
equilibrium is enforced in the SN envelope by conserving photon packets. Level
populations are computed adopting a modified nebular approximation, which
successfully accounts for the deviation from LTE. The code has been successfully
applied to a number of different SNe\,Ib/c \citep[\eg][]{Mazz02ap}, and it has
proved a very useful tool to go beyond line identification: thanks to its
physical treatment of opacities it allows different density and abundance
profiles to be tested, so that various SN properties can be determined
quantitatively. 

Since our code uses a physical luminosity, flux-calibrated spectra are necessary
for a meaningful comparison. The spectra are assumed to be reasonably well
flux-calibrated based on the observational techniques used to obtain them.  We
corrected them for a Milky Way extinction of $E(B-V) = 0.012$ (Schlegel et al.
1998) and measured R-band magnitudes using the PYSYNPHOT
\footnote{http://stsdas.stsci.edu/pysynphot/} package.  In order to account for
calibration errors we applied a large error of 0.3\,mags for both measurements. 
The corrected observed magnitudes are $R = 17.95$\,mag for the Gemini spectrum
and $R= 18.15$\,mag for the Keck spectrum.

Our code requires as input a density distribution. This is essential in order to
define the properties of the SN. In particular, SNe\,Ic with progressively
broader lines are characterised by increasing values of the parameter \KE/\Mej.
Low-energy, ``normal" SNe like SN\,1994I have \KE/\Mej$ \sim 1$ \citep{Sauer06},
while GRB/SNe like SN\,1998bw, which show very broad lines, have \KE/\Mej$ > 2$
\citep{nakamura00}. Given the spectral similarity between SN\,2010ah, SN\,1998bw
and SN\,2002ap, we adopt the explosion model that was used for SN\,1998bw,
rescaling it to match the parameters of SN\,2010ah as estimated above. 

We use a distance of 216\,Mpc (\ie a distance modulus $\mu = 36.67$\,mag)
assuming $H_0=72$\,\kms\,Mpc$^{-1}$ and standard cosmology, and a reddening
$E(B-V) = 0.012$ \citep{corsi11}. 

Given the estimated mass and energy of SN\,2010ah, we started by rescaling the
model for SN\,1998bw \citep[model CO138, \Mej$ = 10 \Msun$, 
\KE\;$= 3\: 10^{52}$\,erg,][]{iwa98} by a factor between 0.3 and 0.5 in mass, 
which obviously implies a reduction in energy of the same magnitude. 

Since both spectra are relatively early, our models can only probe the outer
part of the ejecta. Therefore, while they can yield a reasonable estimate of the
one-dimensional energy, which is mostly associated with the high-velocity outer
ejecta, they cannot probe the bulk of the mass, which is located at lower
velocities and deeper layers than the early spectra can reach. A more accurate
estimate of the mass in this case requires modelling the light curve. Nebular
spectra would be very useful for this purpose \citep[\eg ][]{Mazz07gr}, but none
are available for SN\,2010ah. 

A fundamental element is the choice of risetime. The longer the risetime, the
larger the ejected mass has to be in order to obtain a reasonable spectrum.
SN\,2010ah was first detected on 23.5 Feb 2010. The first spectrum was obtained
on 2 Mar 2010, \ie\ 7 days later.  A deep non-detection limit 4 days prior to
discovery means that the upper limit for the age of the first spectrum is 11
days (10.5 rest-frame days). Averaging these values, we assume $t = 9$\,days as
the epoch of the first spectrum. Considering the redshift $(z = 0.0498)$, we
adopted for the 2 Mar spectrum $t = 8.5$\,d. 

Our code computes a spectrum starting from the SN physical parameters $L, t$,
and $v$, in order to define a consistent temperature structure in the ejecta.
The photospheric velocity $\vph$ can be inferred from the spectra.
\citet{corsi11} measured a \SiII\:6355\,\AA\ absorption velocity $v =
29000$\,\kms\ in the 2 March spectrum. We adopt for the model a slightly lower
velocity, $25000$\,\kms, and obtain a reasonable solution for the spectrum.

\begin{figure}
\includegraphics[width=89mm]{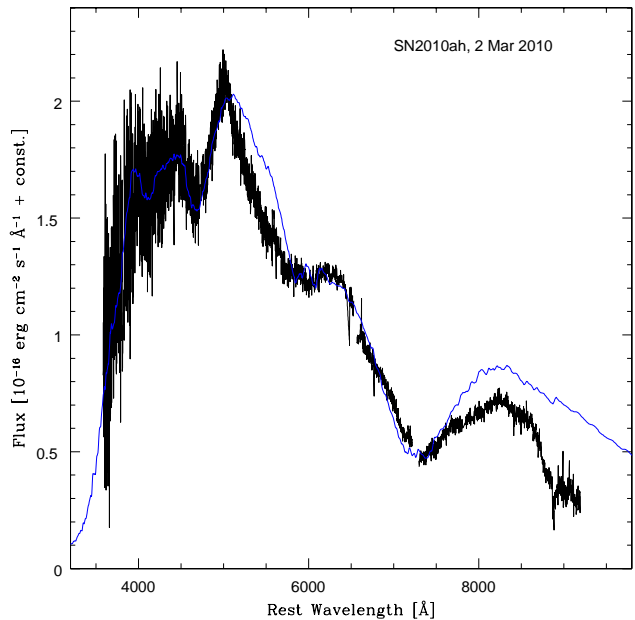}
\caption{The spectrum of SN\,2010ah obtained on 2 Mar 2010 (black) and the 
corresponding synthetic model (blue).}
\label{max}
\end{figure}

The luminosity necessary to reproduce the flux in the 2 Mar spectrum is $5 \;
10^{42}$\,\ergs, \ie $\log(L/\Lsun) = 9.12$. Given the rather small epoch, we
find that the best results are obtained if model CO138 is rescaled in mass by a
factor 0.3. This gives \Mej$\, = 3.2\: \Msun$; \KE$\; = 10^{52}$\,erg, but we
should remark once more that the mass derived from the spectral fits alone can
be highly uncertain since the spectra do not cover a large range of epochs. It
appears however that a much smaller mass is located at high velocities than in
the case of SN\,1998bw, so we need to reduce the mass at $v > 30000$\,\kms. We
did this by steepening the density slope, using a power-law $\rho(r) \propto
r^n$ with index $n \approx -7$ above 20000\,\kms. We find $\approx 0.2 \Msun$ at
$v > 0.1 c$, compared to $\approx 1 \Msun$ for SN\,1998bw. There is however
practically no material at $v > 10^5$\,\kms, as was also the case for
SN\,1998bw.  On 2 Mar 2010, only $\sim 0.5 \Msun$ of material is found above
$\vph$. The equivalent model for SN\,1998bw had $\sim 1.2 \Msun$ located above
$\vph = 31600$\,\kms\ on 3 May 1998 \citep{iwa98}, 8 days after explosion.
Therefore the velocity of SN\,2010ah is actually slightly smaller than that of
SN\,1998bw. 

The synthetic spectrum is compared to the observed one in Fig. 5. Main features
are \FeII\ absorption at $\sim 4000$, 4600\,\AA, \SiII\ near 6000\,\AA, the
\OI-\CaII\ blend near 7000\,\AA. The observed spectrum is reproduced adopting a
composition above the photosphere dominated by oxygen (78\% by number), followed
by Ne (20\%), intermediate-mass elements (5\%). Ca and Fe are only present at
the level of $\sim 0.5$\%. 

\begin{figure}
\includegraphics[width=89mm]{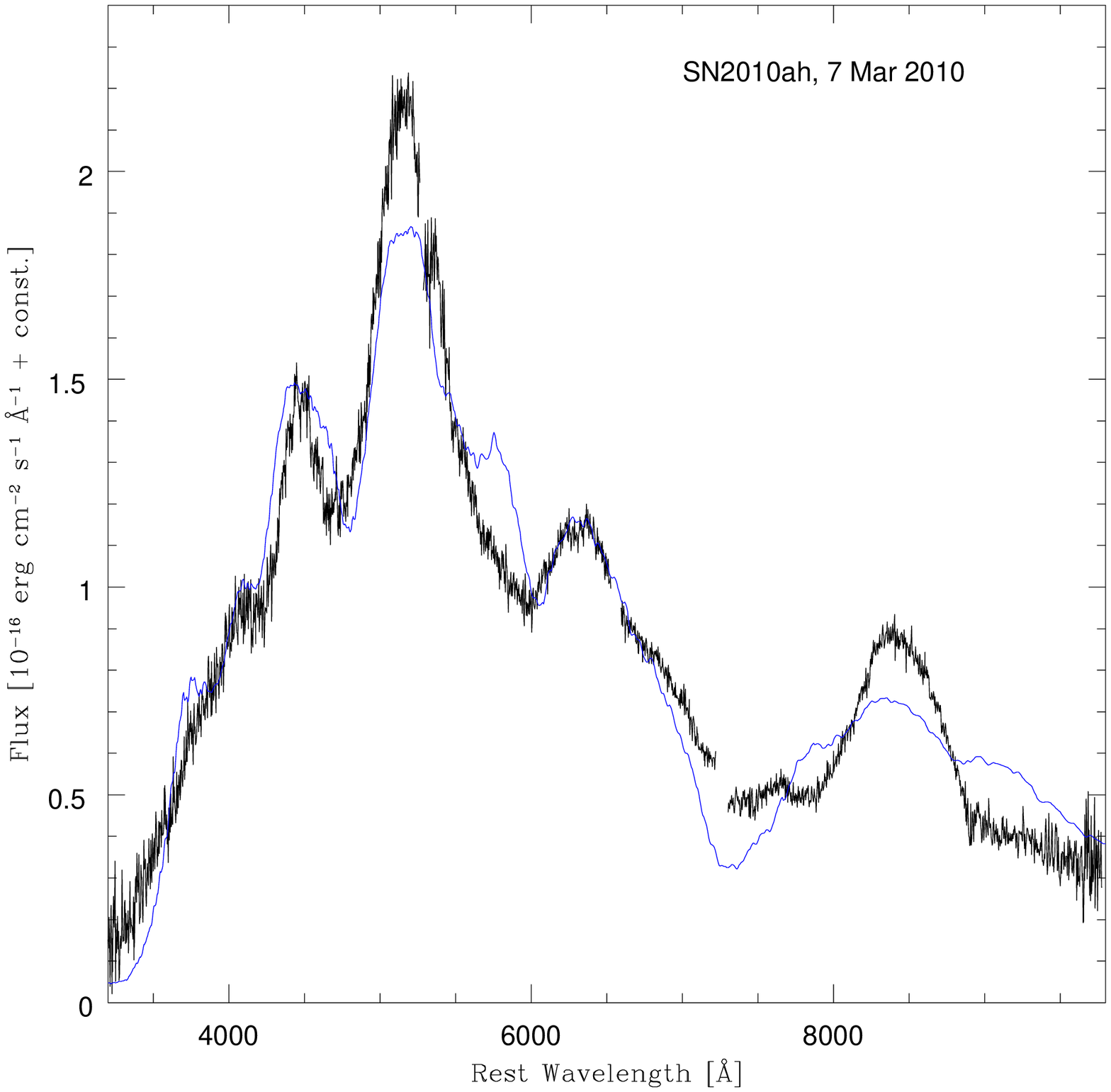}
\caption{The spectrum of SN\,2010ah obtained on 7 Mar 2010 (black) and the 
correspnding synthetic model (blue).}
\label{pl5}
\end{figure}

The next spectrum is only 5 days later, 7 Mar 2010. This corresponds to 14 days,
(13.3 days in the rest frame). The velocity in this spectrum has dropped
significantly \citep[][ measured 18000\,\kms\ in the \SiII\, line]{corsi11},
which is compatible with the drop observed in both GRB/HNe and non-GRB/HNe. We
can thus probe deeper layers of the ejecta. The luminosity required to match the
flux is $L = 4 \; 10^{42}$\,\ergs, \ie $\log(L/\Lsun) = 9.01$. A velocity of
$\vph = 13500$\,\kms\ gives the best fit to the spectrum using the same density
distribution as at the earlier epoch rescaled to the later epoch (Fig. 6). The
photospheric velocity drops by much more than the observed \SiII\ velocity,
suggesting that the observed feature is affected by other contributions. On the
other hand, our synthetic \SiII\ line has a somewhat lower velocity than the
observed one. At this epoch the mass above the photosphere is $2.25 \Msun$. The
composition is similar to that used for the previous epoch, with a little less
oxygen (60\%), replaced by neon (30\%). The spectrum is still formed inside the
oxygen layer. Abundances do not indicate enrichment of Intermediate-mass
Elements or mixing-out of Fe.
The synthetic \OI\ line is too deep, but it is blended with \CaII\,IR, which
actually dominates in strength near 7500\,\AA. A lower density may lead to a
better \OI\ line, but the \SiII\ line would look worse. Since we only have these
two spectra, we refrain from performing a more detailed but uncertain analysis.
We rather turn to the light curve in order to determine parameters such as \Mej,
\Nifs\ mass and distribution.

\section{Light curve models} 

It is well known that light curve models alone do not have the ability to
determine uniquely all SN parameters. In particular, while the mass of \Nifs\
can be reasonably well established from the light curve peak if the time of
explosion is known to within a good accuracy, or from the late, exponential
decline phase of the light curve if this is observed, the two parameters that
determine the shape of the light curve, \Mej\ and \KE, are degenerate 
\citep[Eq.2;][]{Arnett82}. Only the simultaneous use of light curve and spectra 
can break this degeneracy \citep{Mazz97ef}. Having established the most likely
values of \Mej\ and \KE\ through spectral modelling, we can test whether they
can reproduce the light curve \citep{Mazz02ap}. Fitting the light curve yields
additional information about the mass and distribution of \Nifs, in particular
at velocities lower than the photosphere sampled by the spectra. This is
particularly useful for SN\,2010ah, since for this SN we do not have nebular
spectra, which would make it possible to obtain an accurate description of the
inner ejecta \citep[\eg][]{Mazz07gr}. 

The properties of the light curve of a SN of Type I (\ie H-free) are determined
by the mass of \Nifs, which determines the overall luminosity, and by the mass,
density and composition of the ejecta. Mass and energy have a direct impact on
the density of the ejecta as a function of time, and hence on the diffusion time
of the photons. The abundances in the ejecta are important in that they
determine the opacity \citep[in a Type I SN this is dominated by line
opacity;][]{Paul96}. If both early and late-time spectra are available, all
these quantities can be independently determined through spectral modelling.
Typically only small modifications to the mass and distribution of \Nifs\ are
required to reproduce the light curve and validate the results \citep{Mazz03bg}.
This is however not the case for SN\,2010ah, since no nebular spectra are
available. Here we must rely on light curve modelling alone to establish the
mass and density of the inner layers. Still, the properties of the outer layers
are known from the early-time spectra, and so \KE\ is reasonably well
established. 

We use the one-dimensional, monochromatic light curve code described in
\citet{capp97} and later modifications. The code is based on the Montecarlo
method: given a density structure of the ejecta and a \Nifs\ distribution, the
emission and deposition of $\gamma$-rays and positrons produced in the
radioactive decay of \Nifs\ to \Cofs\ and hence to \Fefs\ are computed.
Deposition is assumed to be instantaneous, and to be followed by the emission of
optical photons. Photon diffusion, which determines the light curve, is also
followed with a Montecarlo scheme. Optical opacity is treated assuming that line
opacity dominates: accordingly, it is assumed to depend on composition, as in
\citet{MazzPods06}, reflecting the different number of active spectral lines in
different atomic species. 

\begin{figure}
\includegraphics[width=89mm]{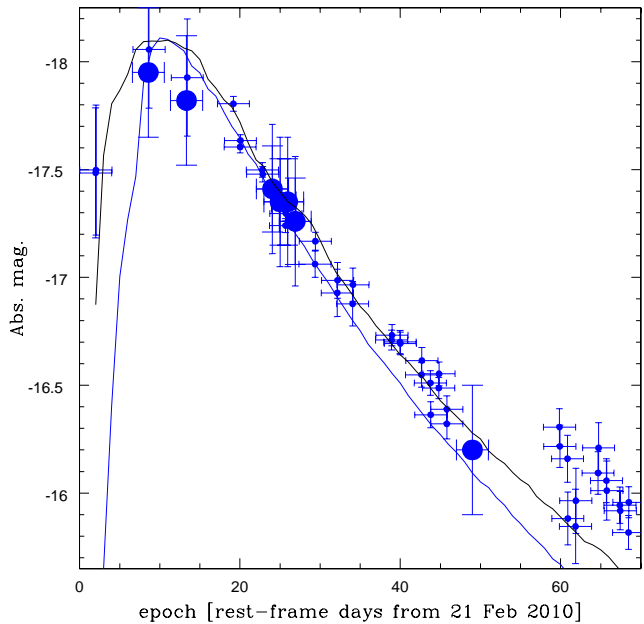}
\caption{The $uvoir$ light curve of SN\,2010ah and two models, one obtained
using the rescaled explosion model used to fit the spectra (blue) and one where
corrections have been applied for mixed-out \Nifs\ and a dense inner core
(black). The bigger blue points are those obtained from direct integration of
the observed magnitudes from the spectra (near peak), from Subaru photometry 
($\sim 25$ days), and at one late epoch (14 April 2010, 49 rest-frame days) 
when multiple photometry was available.}
\label{lcmodel}
\end{figure}

Initially, we use as input the density/abundance profiles derived from
early-time spectral modelling. In particular, the model we used has \Mej$= 3.0
\Msun$, \KE$ = 1.15 \times 10^{52}$\,erg.  The result is the synthetic light
curve shown in blue in Fig.7. This  reproduces the maximum using a \Nifs\ mass
of $0.19 \Msun$, but the rise is delayed relative to the observations. The
decline is also well reproduced, but only until day 30 or so. Neither of these
problems is new in the context of SN\,Ic light curve modelling: the slow rise of
the synthetic light curve is typically the result of insufficient mixing-out of
\Nifs, while the slow observed late-time decline is actually common to all
SNe\,Ic, and a solution was proposed that involves a dense low-velocity core,
possibly a signature of an aspherical explosion \citep{maeda03}. 

We therefore modified both the inner density and the outer \Nifs\ abundance, and
obtained the light curve shown in black in Fig. 7. This modified model has a
power-law index $n \approx -1$ below 20000\,\kms. It also has a slightly larger
mass: $3.3\: \Msun$. The additional $0.2 \Msun$ are located at $v <
10000$\,\kms. The \Nifs\ distribution is also enhanced: outer layers ($v >
20000$\,\kms) contain a \Nifs\ mass fraction of $\sim 10$\%. This is much larger
than the spectral modelling results, which indicate $\sim 1$\%. The \Nifs\ mass
is now $0.25\: \Msun$ (instead of $0.19 \Msun$), but because of the location of
the additional \Nifs\ there is no effect on the peak of the light curve, only on
the rising phase. The highest \Nifs\ abundance is at the lowest velocities, $v <
3000$\,\kms, where a mass fraction of more than 50\% is reached.  However, most
of the \Nifs\ is located at intermediate velocities, with a mass fraction of
$\sim 15$\% at 5000-10000\,\kms\ and $\sim 1$\% at 10000-20000\,\kms.  The
kinetic energy, on the other hand, is only marginally increased (now $1.19 \:
10^{52}$\,erg instead of $1.15 \: 10^{52}$\,erg). The modified model gives a
better match to the overall light curve of SN\,2010ah. Only the last phase ($t >
60$\,d) shows a discrepancy. A further increase in mass at very low velocities
may be a solution, but we refrain from attempting such models because of the
lack of later points to confirm the slower decline. Additionally, although
galaxy subtraction has been applied homogeneously to all images \citep{corsi11},
since the galaxy brightness in $r$ is comparable to that of SN\,2010ah at about
30 days after explosion, the presence of some residual at epochs $\gsim 60$ days
cannot be excluded. This may explain the excess seen in Fig. 7.

In the modified models \Nifs\ is added mostly at high velocity ($>
20000$\,\kms). This requires reaching a \Nifs\ mass fraction of $\sim 0.1$ in
the outer ejecta. Although the \Nifs\ distribution we have used yields a
reasonably good match to the peak of the light curve of SN\,2010ah, it  is
inconsistent with the abundance of \Nifs\ and decay products required for the
spectra. This discrepancy is potentially interesting.  It may indicate that some
\Nifs\ was ejected at high velocity but not along our line of sight, so that it
caused the rapid rise of the light curve \citep{Maeda06} without  leaving an
imprint on the spectra in the form of absorption lines.  

Although this suggestion depends mostly on the first light curve point, this 
interpretation is appealing in view of the properties of highly energetic
SNe\,Ic.  We know from nebular spectroscopy that the distribution of elements is
not spherically symmetric, in the sense that some \Nifs\ and Fe are found at
high velocities, probably in a funnel-like distribution which may track the
position of the GRB jet in some cases \citep{Mazz98bwneb}, while oxygen and
other unburned elements are  ejected more in a plane, likely the equator of the
progenitor. Two-dimensional models suggest this \citep{Maeda02}. There is
further evidence for asphericity in SNe\,Ib/c.  Nebular studies suggest that
even lower-energy SNe can be aspherical, albeit to a lesser degree 
\citep{Maeda08,tauben09,modjaz08,Maurer10a}. It is also particularly interesting
that in a SN\,Ib linked to an X-ray Flash \citep[SN\,2008D,][]{sod08} spectral
analysis suggests a high kinetic energy and the presence of a jet which was
quenched by the helium shell \citep{Mazz08D}, and later data confirmed 
asphericity in the oxygen distribution \citep{Tanaka08}.

The kinetic energy derived for SN\,2010ah is larger than that of SN\,2008D,  
although smaller than in GRB/SNe, and the ejected mass is compatible with that
of SN\,2008D if the He shell in the latter is not considered. An off-axis,
jet-driven SN was proposed for SN\,2008D \citep{Mazz08D}. More energetic SNe\,Ic
are all linked to GRBs and aspherical \citep[\eg][]{Mazz98bwneb}. Given the
peculiarly high first point in the light curve of SN\,2010ah, an aspherical
distribution of elements may also be speculated for SN\,2010ah.  The lack of
nebular or spectropolarimetric data prevents us from testing this hypothesis.

\section{Discussion}

The results of the analysis we performed can be used to estimate the properties
of the  progenitor of SN\,2010ah.  This is interesting, because of the suggested
relation between \KE, M(\Nifs), and \Mej, which should be a good proxy for the
mass of the progenitor star \citep{Mazz03bg}.  In the case of SN\,2010ah, we
found \Mej$\, \sim 3.3\; \Msun$, \KE\,$\sim 1.2 \; 10^{52}$\,erg (in spherical
symmetry), M(\Nifs)$ \sim 0.25 \Msun$. 

If we plot the properties of SN\,2010ah and those of other well studied SNe
(Figures 8 and 9), we see a trend between \Mej, M(\Nifs), and \KE. Figure 8
shows M(\Nifs) v. \KE. The trend is clear, although SN\,2006aj deviates
somewhat. In this case the birth of a magnetar was suggested \citep{Mazz06aj},
and the energy released may have led to an increased synthesis of \Nifs.
Slightly more massive progenitors probably collapsed to a black hole, thus
reducing the amount of material available for nucleosynthesis. At larger masses
(and larger energies) the efficiency of production and ejection of \Nifs\
increases again. Figure 9 shows M(\Nifs) v. \KE. Here the relation is less well
defined, but a trend is still visible. SNe\,Ib and IIb have a massive He
envelope and lie slightly higher in the plot, suggesting that the \KE\ is not
influenced by the properties of the outer layers, as is reasonable to expect. 

\begin{figure}
\includegraphics[width=92mm]{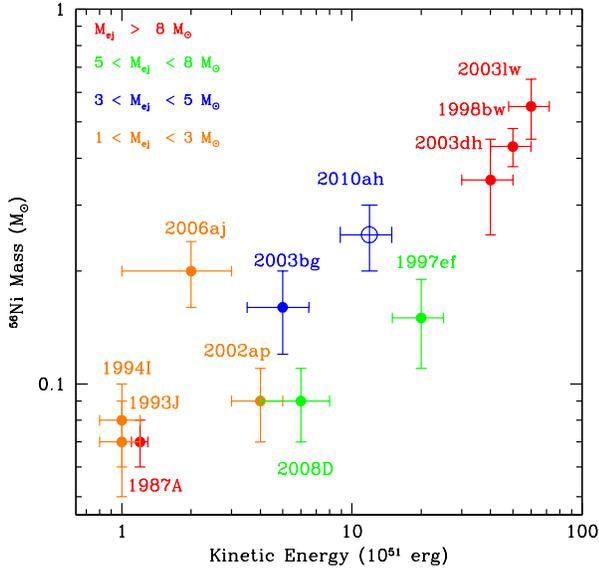}
\caption{The mass of \Nifs\ and the \KE\ as derived from spectral/light curve
modelling of a number of SNe\,Ib/c. Colour-coding is by ejected mass.}
\label{nike}
\end{figure}

\begin{figure}
\includegraphics[width=92mm]{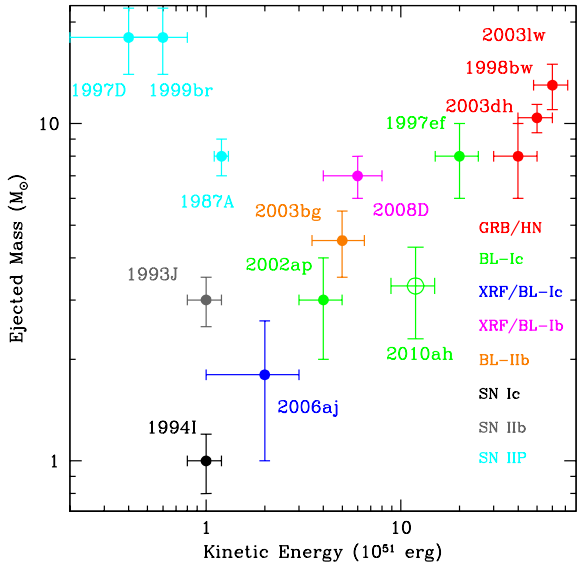}
\caption{The ejecta mass and the \KE\ derived from spectral/light curve 
modelling of a number of SNe\,Ib/c and some SNe\,IIP. Colour-coding is by 
spectral type. }
\label{mke}
\end{figure}

\begin{figure}
\includegraphics[width=92mm]{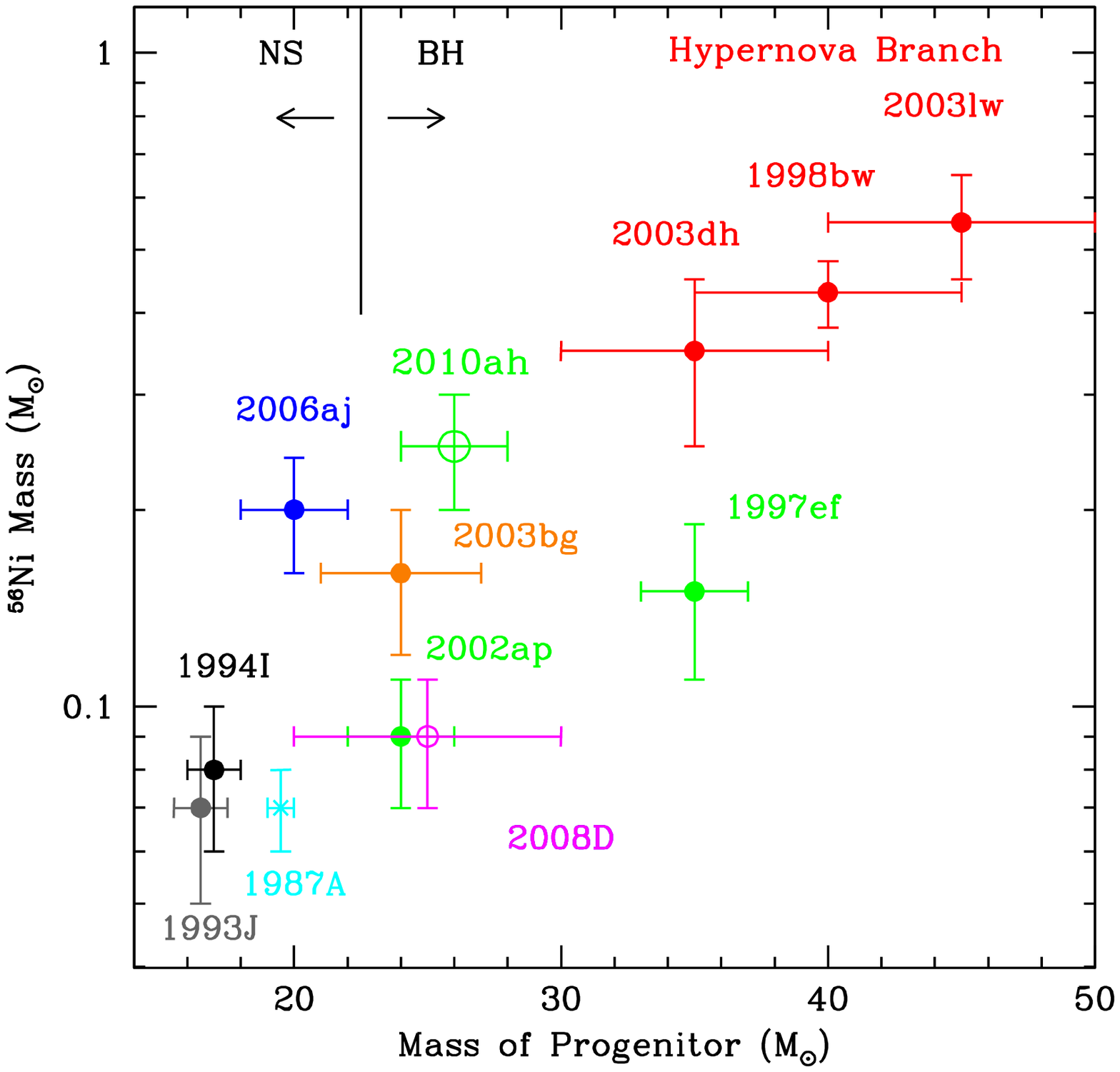}
\caption{The mass of \Nifs\ derived from spectral/light curve modelling of a 
number of SNe\,Ib/c plotted against the inferred ZAMS mass of the progenitor 
star using single-star evolutionary models. Colour-coding is by spectral type,
and it is the same as in Fig. 9.}
\label{nizams}
\end{figure}

\begin{figure}
\includegraphics[width=92mm]{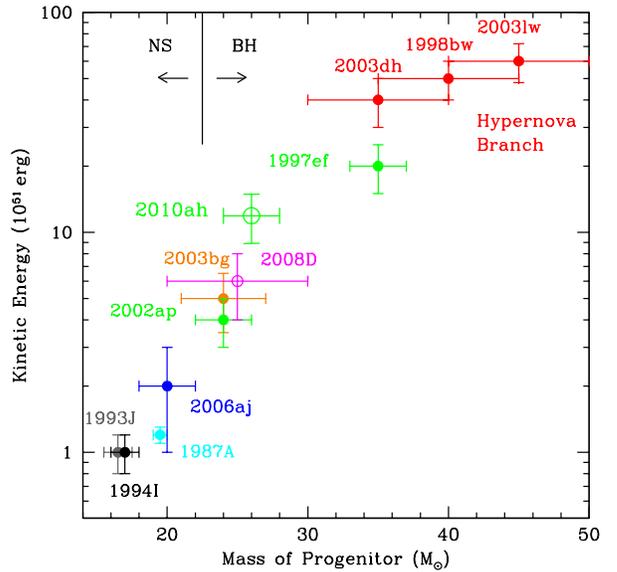}
\caption{The \KE\ derived from spectral/light curve modelling of a number of 
SNe\,Ib/c plotted against the inferred ZAMS mass of the progenitor star using 
single-star evolutionary models. Colour-coding is by spectral type,
and it is the same as in Fig. 9.}
\label{kezams}
\end{figure}

We can try to go from the estimated \Mej\ to a progenitor mass. Using the 
\citet{nomhash88} models, and assuming that the mass ejected by SN\,2010ah
corresponds to the CO core of a massive star minus the remnant mass, we end up
with a black hole remnant, a CO core of $\sim 5-6 \Msun$, and a M$_{ZAMS} \sim
24-28 \Msun$.  In the more traditional plots of M(\Nifs) or \KE\ v. M$_{ZAMS}$
(Figures 10 and 11), SN\,2010ah falls nicely on the relation that is taking
shape \citep[e.g.][]{Mazz03bg}. This relation is derived for single-star
evolution, and using a single, albeit old set of stellar models.  The process of
stripping the outer H and He layers may require interaction with a companion in
a binary system \citep{Fryer07} and may not be easy to parametrise. Also, the
uncertainty on the nature of the remnant (a Neutron Star or a Black Hole?) may
introduce errors in the mass estimate.

Still, in the case of SN\,2010ah, it looks as if all properties we derived
(\Mej, \KE, M(\Nifs)) are consistently smaller than those of GRB/SNe, although
they are large compared to other non-GRB/HNe. We may be testing the limits for
stars to produce GRBs. Another recent non-GRB/HN,
SN\,2010ay \citep{Sanders12} may come even closer to this limit and it will be
an interesting case study. 

A question remains concerning SN\,2010ah: did it produce a GRB, or was it at
least driven by a jet which failed to escape from the star, as was suggested for
SN\,2008D \citep{Mazz08D}? No detection of a GRB has been reported. Off-axis
GRBs may be detected in the radio, but this is not the case for SN\,2010ah
\citep{corsi11}.  The anomalously rapid rise of the light curve may suggest the
presence of highly processed material in a direction not along our line of
sight, but this cannot be regarded as a strong argument in the absence of other
supporting evidence. Spectropolarimetry in the early phase or late-time
spectroscopy would be useful in order to reveal asphericities. These may in turn
point to the presence of a jet which, as in the case of SN\,2008D
\citep{Tanaka08}, may have aided exploding the star even though it may not have
emerged from it. Unfortunately, no such data are available for SN\,2010ah, so we
cannot determine the degree of asphericity or the presence of any asymmetry. A
suggestion for asphericity drawn from the rapid rise of the light curve may be a
point in favour of this. It would be important that exceptional SNe like
SN\,2010ah are studied in more detail in the future.


\section*{Acknowledgments} 
We acknowledge financial support from grants: INAF PRIN 2009 and 2011,
ASI I/016/07/0, ASI I/088/06/0.



\bsp

\label{lastpage}

\end{document}